\documentclass[letterpaper,twocolumn,10pt]{article}
\usepackage{usenix2019, endnotes}
\usepackage{color}
\usepackage{graphicx}
\usepackage{wrapfig}
\usepackage{multirow}
\usepackage{listings}
\usepackage{url}
\usepackage{dblfloatfix}
\usepackage{fixltx2e}
\usepackage{balance}
\usepackage{listings}

\newif\ifdraft
\draftfalse                                                                              
\ifdraft
\newcommand{\zhao}[1]{{\textcolor{cyan}    { ***Zhao:      #1 }}}
\newcommand{\outline}[1]{{\textcolor{blue}    { ***Outline:      #1 }}}
\newcommand{\lei}[1]{{\textcolor{red}    { ***Lei:      #1 }}}
\newcommand{\note}[1]{ {\textcolor{red}    {\bf #1 }}}
\else
\newcommand{\zhao}[1]{}
\newcommand{\outline}[1]{}
\newcommand{\lei}[1]{}
\newcommand{\note}[1]{}
\fi

\newenvironment{shortlist}{
        \vspace*{-0.5em}
  \begin{itemize}
  \setlength{\itemsep}{-0.1em}
}{
  \end{itemize}
        \vspace*{-0.5em}
}

\begin{document}

\title{\Large \bf FanStore: Enabling Efficient and Scalable I/O for Distributed Deep Learning}

\author{
{\rm Zhao Zhang$^1$}\\
\and
{\rm Lei Huang$^1$}\\
\and
{\rm Uri Manor$^2$}\\
\and
{\rm Linjing Fang$^2$}\\
\and
{\rm Gabriele Merlo$^3$}\\
\and
{\rm Craig Michoski$^3$}\\
\and
{\rm John Cazes$^1$}\\
\and
{\rm Niall Gaffney$^1$}\\
\and
{\rm Texas Advanced Computing Center, University of Texas at Austin$^1$}\\
{\rm Salk Institute$^2$}\\
{\rm The Institute for Computational Engineering and Sciences, University of Texas at Austin$^3$}
} 

\maketitle

\subsection*{Abstract}
Emerging Deep Learning (DL) applications introduce heavy I/O workloads on computer clusters.
The inherent long lasting, repeated, and random file access pattern can easily saturate the metadata and data service and negatively impact other users.
In this paper, we present FanStore, a transient runtime file system that optimizes DL I/O on existing hardware/software stacks.
FanStore distributes datasets to the local storage of compute nodes, and maintains a global namespace.
With the techniques of system call interception, distributed metadata management, and generic data compression, FanStore provides a POSIX-compliant interface with native hardware throughput in an efficient and scalable manner. 
Users do not have to make intrusive code changes to use FanStore and take advantage of the optimized I/O.
Our experiments with benchmarks and real applications show that FanStore can scale DL training to 512 compute nodes with over 90\% scaling efficiency.

\section{Introduction}
We have seen an increasing demand of computer cluster resources for commercial and scientific research that are leveraging the Deep Learning (DL) technology. 
Using DL, fascinating results have been obtained in domains such as autonomous driving~\cite{Reiley2016}, the GO board game~\cite{Silver2016}, the checkout-free grocery store~\cite{Alba2016}, astronomy~\cite{Zhang2017GAN}, drug discovery~\cite{Aliper2016}, high energy physics detector design~\cite{Vallecorsa2018CERN}, and bioinformatics~\cite{Plis2013, Chicco2014, Lena2012}. 
A recent survey~\cite{Hoefler2018DDL} shows that over 50\% of 227 published deep learning works are running on multiple nodes to reduce experiment turnaround time.
Meanwhile, researchers have scaled DL training to thousands of compute nodes without losing test accuracy~\cite{You2018, Codreanu2017, Akiba2017}.

Distributed DL training often involves large datasets.
Taking image classification as an example of an I/O intensive DL problem, the ImageNet-1k dataset~\cite{Deng2009imagenet} contains 1.3 million small files (from bytes to megabytes) that are spread across 2,002 directories. 
Training a ResNet-50~\cite{he2016deep} neural network usually runs for 90 epochs, which means every file will be accessed 90 times. The total 117 million file accesses are distributed across the training procedure.
Depending on the scale, the concurrency of file access can be in the order of $O(N)$, where N is the number of processing elements, e.g., CPUs and GPUs.

One way to run distributed DL training on existing cluster hardware and software stacks is to place the dataset in the shared file system and the program access files through the POSIX interface. 
Users usually start building their applications in this straightforward way.
As the dataset grows larger, the metadata and data traffic of thousands of directories and millions of files can easily saturate the existing shared file system due to the high access frequency, concurrency, and the sustained I/O behavior.
In a typical cluster setting, such an I/O workload negatively affects other users with degraded file system performance or even unresponsiveness.
A second way is to encapsulate the larger number of small files in an optimized data format. 
For example, Caffe~\cite{Jia2014} can read data from LMDB~\cite{Chu2011mdb} format, while TensorFlow~\cite{Abadi2016} and MXNet~\cite{Chen2015} have their native TFRecord and IORecord format, respectively.
Using the customized data format can dramatically reduce the metadata workload, but the data workload remains the same and the limited bandwidth between file system and compute nodes is a performance bottleneck.
Additionally, users have to make appropriate code changes to read data from the corresponding format.
Another technical workaround, copying the complete dataset to the local disks of each compute node, is a viable solution with one condition: the local disk is large enough. 
However, many datasets exceed the local storage space.

In this paper, we present FanStore, a runtime shared file system to enable efficient and scalable distributed DL training.
FanStore is designed based on the findings of a profile study on distributed DL I/O behavior (Please find details in \S\ref{sec:profile}).
FanStore exposes a POSIX-compliant interface with the relaxed multi-read single-write consistency, 
so users do not have to make intrusive code changes to take advantage of the optimized I/O performance.
FanStore employs the function interception technique to enable the POSIX-compliant interface in user space.  With little overhead, I/O performance through FanStore is close to the hardware limit.
FanStore preserve the global view of dataset by broadcasting metadata and distributing file data, with remote file access as a round-trip MPI~\cite{Gropp1996mpi} message.
By combining the relaxed consistency and various techniques, FanStore can scale distributed DL training to 512 nodes with over ~90\% efficiency.

This paper makes the following contributions:
\begin{shortlist}
\item{A distributed DL training I/O profile study, with identified opportunities for file system consistency relaxing and I/O optimizations.}
\item{The investigation of various techniques that make high utilization of existing hardware with improved overall training performance.}
\item{The verification of the effectiveness of FanStore's design using three real applications that cover the convolutional, recurrent, and generative adversarial network architectures.}
\item{The open source implementation of FanStore.}\endnote{Code will be released on GitHub later.}
\end{shortlist}

The rest of this paper is organized as following:
The motivating applications and DL I/O profile study is presented in \S\ref{sec:apps} and \S\ref{sec:profile}.
We summarize related work with comparison to FanStore in \S\ref{sec:related}.
\S\ref{sec:design} discusses the FanStore design and implementation.
Performance measurements with benchmark and real applications are presented and discussed in \S\ref{sec:experiments}
Then we draw conclusion and envision future work in \S\ref{sec:conc} and \S\ref{sec:future}.

\section{Motivating Applications}
\label{sec:apps}
FanStore is motivated by real world applications.
Examples include ResNet-50, super resolution generative adversarial network (SRGAN), and Fusion Recurrent Neural Network (FRNN). 

ResNet-50 is a 50-layer convolutional neural network for image classification.
The test case is training ResNet-50 with the ImageNet-1k dataset, which has 2,002 categories and 1.3 million images in total.

SRGAN~\cite{ledig2017photo} in this case is processing 3D scanning electron microscope imaging data of neural tissue samples. It trains a generative adversarial network (GAN) to increase the resolution of undersampled images acquired on a point scanning imaging system (e.g. a scanning electron or scanning confocal microscope). By training on image pairs acquired at full resolution and 4x undersampled resolution, the network can effectively perform “compressed sensing” without any of the typical pitfalls, including specialized acquisition software, reconstruction artifacts, and slow computational speeds. Most importantly, when imaging 4x undersampling on a point scanning imaging system, the imaging speed is effectively increased by 16x, while the potentially damaging photon or electron dose is decreased by 16x. Hence the ability to upscale undersampled images with a neural network provides significant advantages at both the pre- and post-processing stages of the imaging experiment.
SRGAN has two stages: initialization and training. 
The example code runs 100 initialization epochs and 2000 training epochs.
We report the performance of both parts in subsequent results.

FRNN is to predict disruptions in tokamak reactors. 
It dramatically improves/optimizes both the performance and operating/repair costs of these reactors. 
Practical disruption prediction in tokamaks has recently improved by utilizing
deep learning algorithms with real-time machine diagnostics to predict via long short term memory (LSTM) the onset of major disruptions.   
FRNN enhances these algorithms by incorporating and propagating experimental
uncertainties from diagnostic signals.
The dataset comes with 171,264 files in one directory. The total size is $\sim54$~GB. 
~\zhao{put dataset stats once it is ready}

Table~\ref{tb:apps} presents the neural network architecture (NN Arch), the underlying DL framework and distributed DL method.
While Table~\ref{tb:data} summarizes the statistics of the three datasets.

\begin{table}[t]
\begin{center}
    \caption{Summary of Applications}
    \begin{scriptsize}
    \begin{tabular}{ | c | c | c | c |}
    \hline
    App & NN Arch & DL software & Distributed method  \\ \hline \hline
    ResNet-50 & CNN & Keras+TensorFlow & Horovod \\ \hline
    SRGAN & GAN  & TensorLayer+TensorFlow & Horovod \\ \hline
    FRNN & RNN & TensorFlow & MPI4py~\cite{dalcin2005mpi}\\ \hline
    \end{tabular}
    \end{scriptsize}
    \label{tb:apps}
\end{center}   
\end{table}

\begin{table}[t]
\begin{center}
    \caption{Basic Statistics of the Datasets}
    \begin{scriptsize}
    \begin{tabular}{ | c | c | c | c | c |}
    \hline
    App & \# files & \# dirs & total\_size & file\_size \\ \hline \hline
    ResNet-50 & 1.3 million & 2,002 & 140~GB & KB-MB \\ \hline
    SRGAN & 0.6 million & 6 & 500~GB & MB \\ \hline
    FRNN & 0.17 million & 1 & 54~GB & KB\\ \hline
    \end{tabular}
    \end{scriptsize}
    \label{tb:data}
\end{center}   
\end{table}

\section{DL I/O Profile}
\label{sec:profile}
In this section, we first review how distributed DL works in a data parallel manner and the corresponding I/O behavior. 
Then we summarize the I/O frequency, concurrency, and consistency.
For the ease of understanding, we will use the ResNet-50 training case with ImageNet-1k dataset implemented with Keras~\cite{Chollet2015keras}, TensorFlow~\cite{Abadi2016}, and Horovod~\cite{sergeev2018horovod}.
Keras provides a concise programming interface with back end support of TensorFlow and other frameworks, 
while Horovod works at the communication layer of TensorFlow and enables distributed training.
The summarized I/O pattern applies to the three example applications discussed in this paper and many other applications too.

\subsection{Data Parallel Distributed DL}
Data parallel is a common way to enable distributed DL training, as it can efficiently use the computation resources.
In the data parallel approach, the model (parameters) is replicated on all nodes.
In the very beginning of training, the program will traverse the metadata in the training and testing directories to calculate the number of files, then determine the number of iterations in each epoch and in total.
From the first iteration, each node will concurrently read a mini-batch of training files.
The size of the mini-batch is a user specified parameter, and the mini-batch size is critical to final convergence: mini-batch sizes larger than published numbers are training divergence prone.
The training process then carries out forward computation along the neural network, and computes the loss in the end. 
After that, each node will use the computed loss to calculate the gradients with regard to parameters, which is referred to as a back-propagation algorithm.
Since each node has a different set of training items, the derived loss and respective gradients are different across nodes.
Usually, the training process calls the Allreduce collective communication primitive to compute the sum (then mean) of the gradients before applying the stochastic gradient descent (SGD) method to update parameters.
One iteration finishes after all parameters are updated. 
If the end of the current iteration overlaps with the end of an epoch, the training process may validate the model on the test dataset and checkpoint the model to file system.
The complete training process iterates until all epochs finish.

\subsection{Global Dataset View}
\begin{figure}[h]
	\includegraphics[width=85mm]{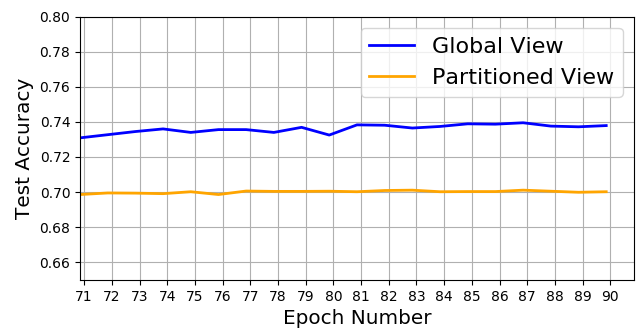}
	\caption{Test Accuracy of ResNet-50 on ImageNet-1k Dataset with Global and Partitioned View}
 	\label{fig:divergence}
\end{figure}

With the global dataset view, every compute node sees the same directory structure and file contents.
Maintaining a global view of the training dataset is critical for convergence in distributed DL training. 
Another way to store a large dataset in local storage across multiple nodes is to let each node store an exclusive subset, which results in the partitioned dataset view.
Figure~\ref{fig:divergence} shows the last 30 epochs of the 90-epoch ResNet-50 training on the ImageNet-1k dataset with both views.
The partitioned dataset view loses $\sim4$\% of test accuracy, which is unacceptable.

\subsection{Metadata Access}
\label{subsec:metadata}
At a high level, the metadata access of distributed DL has high volume and concurrency.
Metadata is accessed in two places.  One is at the beginning of the training process, where the programs need the information of the training and test dataset. 
The other is during each iteration, where the training program launches multiple threads per process to read files.
For example, each Keras process uses four I/O threads by default.

In the ImageNet-1k dataset, each process accesses the metadata of 2,002 directories and 1.3~million files at the beginning of training. 
On a cluster of $N$ GPUs, where we usually run one process per GPU, there will be $4N$ simultaneous {\it readdir()} or {\it stat()} operations.
The highly concurrent metadata access with large volume can easily saturate the metadata server in a traditional shared file system, where there may be only one single metadata server such as Lustre~\cite{schwan2003lustre} or enforced locking to guarantee consistency such as GPFS~\cite{schmuck2002gpfs}.

\subsection{Data Access}
\label{subsec:data}
The file data access pattern of distributed DL features high concurrency and such a pattern persists through the whole training process.
The individual file size ranges from a few bytes to a few mega bytes.
Modern DL frameworks such as Keras and Caffe supports asynchronous I/O, where the I/O overlaps with computation for faster training speed. 
In some cases, this is referred to as data prefetch.
Assuming a cluster of $N$ GPUs, the mini-batch size is specified proportionally to GPU count to maintain high utilization of the hardware.
In the ResNet-50 example, we use $64N$ mini-batch size.
Thus the data access is in the form of $4N$ concurrent threads reading $64N$ files for each iteration. 
One iteration of the ResNet-50 example runs for $\sim500$~ms.  If the I/O performance cannot keep up with the pace of computation, there will be wasted hardware cycles.
Such a data access pattern persists until the very last iteration.
It is not rare to use tens, hundreds, or even thousands of GPUs or CPUs.
Thus for distributed DL training, it is critical for the file system to keep up with the computational hardware in a scalable manner.

When a file is read, it is read sequentially and completely. There is no random read that starts from an arbitrary offset nor partial read from a file.

Besides accessing training and test dataset, 
distributed DL writes to file system in some cases.
The master process periodically writes the model to file system as a checkpoint.
In applications such as generative adversarial networks, the training program may output the generated synthetic data to file system for human examination.
All the write operations are writing to new files without overwriting or concurrent writing to the same file.
\endnote{Though the checkpoint can be overwritten, writing to a file labeled with epoch number is a commonly adopted practice.}
Unless the training program starts from the last checkpoint, these written files are not read again by the training program.

\subsection{Consistency}
The I/O behavior of distributed DL training is a multi-read single-write consistency pattern.
A directory or file can be accessed by multiple processes/threads simultaneously, while each output file is written exclusively by a single process/thread with no further access.
Thus it is free of read-after-write or write-after-write hazards.

\section{Related Work}
\label{sec:related}
Massively concurrent access of small files has been studied extensively by researchers.
This section summarizes previous work on the optimization on file systems and the I/O systems, with an analysis of their applicability in distributed DL training.

Using cachefilesd~\cite{cachefilesd2018} can cache datasets from a shared file system to local storage with a FIFO eviction policy. This method can indeed enhance I/O performance if the dataset is small enough. 
Also it lacks supports for massively concurrent metadata access. 

Since the emergence of data intensive computing on supercomputers~\cite{Wilde2009swift, Raicu2008falkon}, researchers have noticed that the conventional single metadata server design in a shared file system can not accommodate massive small file I/O operations efficiently~\cite{Zhang2013envelope}. 
Various file system optimizations have been proposed.

One type of optimizations focuses on distributed metadata server design.
GIGA+~\cite{Patil2011giga} contributes a dynamic metadata server scaling design for incremental file count growth. 
While ZHT~\cite{Li2013zht} and GlusterFS~\cite{Davies2013glusterfs} propose a static zero-hop hash table for scalable metadata management.
Both metadata server designs can achieve decent scaling performance for massively concurrent file I/O. 
Since the metadata is spread across servers, directory access with {\it readdir()} has to communicate with all servers to gather information.
Thus both designs will result in slow directory access for distributed DL training, given the large metadata volume and highly concurrent access.

A second way of optimization seeks to relax file system consistency.
HDFS~\cite{Shvachko2010hdfs} only supports file appending to achieve high output throughput.
A multi-read single-write consistency is implemented in AMFS~\cite{Zhang2012} to provide highly efficient and scalable I/O support for workflow applications on supercomputers.
In the context of distributed DL training, further relaxing file system consistency can achieve better performance as the output files are never read by the training program.

Users usually have to mount these file systems through the FUSE~\cite{FUSE} interface, due to the lacking of root privileges. 
The overhead through FUSE is not trivial~\cite{Vangoor2017fuse}. Thus the I/O performance is sub-optimal compared to the kernel module case.
In our work, we use the system call interception technique to remedy this performance issue.

Optimization of massive small file access can also be applied with customized data formats.
Using high performance data formats such as HDF5~\cite{Folk1999hdf5} can potentially achieve higher metadata access performance, and each file access will be in the form of record retrieval.
Major DL frameworks work with datasets stored in customized formats. 
E.g., Caffe works with LMDB, while TensorFlow has its own TFrecord format, and MXNet has IOrecord.
The problem with these data formats is that users have to make intrusive code changes to the application if they start with POSIX-compliant files.
Despite the high performance and scalability, FanStore preserves the POSIX-compliant interface for users to transition from a small dataset to a large dataset without additional code changes.

\section{FanStore Design and Implementation}
\label{sec:design}
In this section, we discuss the FanStore architecture and implementation details.

\subsection{Overview}
\begin{figure}[h]
	\includegraphics[width=85mm]{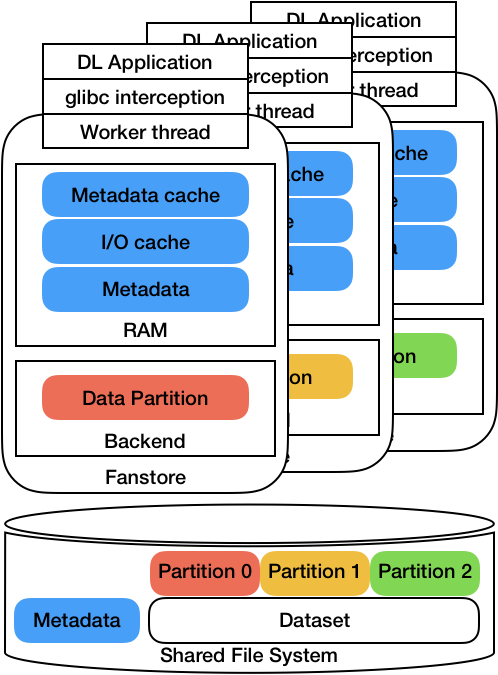}
	\caption{FanStore Architecture Overview}
 	\label{fig:arch}
\end{figure}

Figure~\ref{fig:arch} shows the the overview of FanStore architecture.
FanStore places metadata and file data in RAM and local disks, respectively. 
It maintains several data structures for high throughput file access.
One or more worker threads within each FanStore process handle file system
requests intercepted from the DL training process.
These worker threads manipulate the metadata stored locally and retrieve file data either from local storage or remote node via network.

\subsection{Data Preparation}
FanStore requires a data preparation step before training, 
where a user will have to pass into a preparation program a list of all files involved.
Large datasets originally stored in the shared file system are then reorganized into partitions.
Each partition contains an exclusive subset of the files.
Table~\ref{tb:layout} shows the data layout in a partition.
Each partition starts with an integer (eight bytes) of the file count, followed by a 256 byte long file name, a 144 byte long stat structure as the file's metadata, and the data size after compression.
Then the actual data is appended. 
The rest of the files are organized continuously.
If the data is not compressed, the actual data size is in the stat structure and compressed\_size is set to zero.
Otherwise, the actual data size is stored in compressed\_size.

Upon loading, FanStore traverses each partition to dump the actual data into local storage and builds an index of file path and storage place, which includes both the node id and the data offset.
Such a design dramatically reduces the metadata count compared to the case of storing the files on the shared file system.

\begin{table*}[t]
\begin{center}
    \caption{Data Layout in a Partition}
    \begin{tabular}{ | c | c | c | c | c | c | c | c |}
    \hline
    field & num\_files & file\_name & stat & compressed\_size & data & file\_name & stat \\ \hline \hline
    byte\_range & 0 - 3 & 4 - 259 & 260 - 403 & 404 - 411 & 412 - 411+data.size & ... & ... \\ \hline
    \end{tabular}
    \label{tb:layout}
\end{center}   
\end{table*}

When using FanStore, it prefixes the original relative file path with a predefined mount point. For example, the path {\it ILSVRC2012\_img\_train} on the shared file system will be available as {\it /fanstore/user\_id/ILSVRC2012\_img\_train}.
The internal structure remains unchanged.
All nodes share the same view of the namespace of training and test datasets.

\subsection{Metadata Management}
FanStore keeps metadata in a hashtable in RAM. 
Each entry has the file path as the key and the metadata record as the value.
Besides the POSIX-compliant information, each metadata record maintains the file location. 

As discussed in \S\ref{subsec:metadata}, efficiently and scalable DL training requires high throughput metadata access, especially given the large volume and highly concurrent I/O traffic.
The input files of the training and test datasets remain unchanged during training, while the output files are written only once. 
Thus, FanStore employs different strategies for input and output files.
All the metadata of input files is replicated across nodes. This replication provides each node with an identical view of the metadata.
In each FanStore process, the file metadata of a directory is preprocessed and cached in a hash table to allow {\it readdir()} to return immediately.

Overall, the metadata of output files is distributed across all nodes using a consistent hash function.
A particular file maps to a node using the modulo of the path hash value and the node count.
The metadata of an output file has only one copy on the mapped node.
The output file metadata is not visible until all write operations finish and the file descriptor is closed.

\subsection{Data management}
\outline{Different access patterns for training and test datasets.}
We have observed two access patterns for training and test datasets:
the training dataset is usually larger than the test dataset, and each training process can randomly access files in the training dataset during each iteration.
In contrast, the test dataset is completely read by each process during validation, and this is usually done at the end of each epoch.
Based on this observation, FanStore allows users to specify a directory so that all files in this directory will be replicated across all nodes.
This replication can improve validation performance due to higher locality hit rate.

\outline{Striping}
As discussed in \S\ref{subsec:data}, an input file is read completely in a sequential order. 
So, FanStore stores each input file as a byte array without block abstraction or striping.

\outline{Read request}
Upon receiving a file open request, the worker thread checks its availability and location in metadata.
If the file exists in local storage, the thread pulls the file from local storage to memory then returns the file content; if the file exists on a remote node, the thread communicates with the peer thread on that node to retrieve the file content; if the file does not exist, it returns an error code.
The communication in FanStore is implemented using MPI for high bandwidth and low latency.

\outline{Caching}
Each file in the training dataset has a uniform probability of being accessed, and the access of each file is independent of other file access operations.  
It is difficult to design a cache eviction policy for high hit rate.
FanStore implements an easier caching mechanism: a file is cached in memory until the file descriptor is released. 
We intend to use as little RAM space as possible given that the training process can be memory hungry.
Occasionally, multiple training processes on the same node can access the same file simultaneously. 
Closing the the file descriptor or evicting the file from cache can result in a stale state in other process.
FanStore maintains a file counter table in memory with file path as the key and the number of processes that are currently accessing it as the value.
Once a file is accessed, the corresponding counter increases by one. 
Upon the release of a file descriptor, the corresponding counter decreases by one.
If the counter is zero, the file content is evicted from cache.

\outline{Compression}
To expand the capacity of existing storage hardware, FanStore can compress the file data during dataset preparation.
The input file access thus needs one more step of decompression before returning file content.
We use the LZSSE8 library~\cite{LZSSE} that implements the Lempel-–Ziv–-Storer–-Szymanski lossless compression algorithm~\cite{Storer1982data}.
LZSSE8 allows various compression levels as a tradeoff between compression speed and ratio.

\outline{Replication}
\zhao{replication, there is another case where the node count $>$ partition count}
When scaling out distributed DL training, there will be more aggregated local storage space, though the dataset may not fit in a single node. 
In this case, FanStore allows users to specify a replication factor of N, so that each node can host N different partitions.

\outline{Write request}
For the writing case, FanStore implements a visible-until-finish consistency. 
The newly created metadata is temporarily kept on the originating node and the data written is concatenated to a buffer. 
Once all writes finish and FanStore receives the {\it close()} request, the metadata entry will be forwarded to the corresponding node mapped by the consistency hash function.

\subsection{POSIX Interface}
Users on clusters usually do not have root privilege, which makes it infeasible to mount FanStore as a kernel module. 
For many other file system implementations, such as GlusterFS and PVFS, exposing POSIX-compliant interface through FUSE is a viable solution.
However, FUSE introduces non-trivial overhead as the system call crosses the user-kernel boundary.
Such overhead results in significant slowdown in DL training.

To eliminate the performance overhead while preserving user-space usability, FanStore implements the POSIX interface using the function interception method~\cite{hunt1999detours}.
I/O operations from applications eventually call the low level functions such as {\it open(), close(), stat(), read(), write()} in the GNU C Library (glibc).
The function interception method replaces the first several instructions of the low level functions in glibc and forces them to jump into a user space library where FanStore logic is implemented. 
In this way, all I/O related function calls stay in user space.
Our current implementation only supports x86 architecture.
We present the performance improvement from this method in \S\ref{subsec:benchmark}.

\subsection{Resilience}
FanStore does not address resilience in system design.
A process failure or a node failure is usually accompanied with the unavailability of computing resources.
In that case, distributed DL training has to halt until the node recovers, as losing computing resources results in a smaller batch size, which is very sensitive to test accuracy. 
This is particularly important for the synchronous stochastic gradient descent method.
Any parameter updates with partial results can lead to a stale model state.
On the application side, periodically checkpointing model to the shared file system is a commonly adopted approach, as it is natural for a previous model to have higher test accuracy than the later ones, and users will take the previous model as training results.
Only one process writes the checkpoint in most cases, as the model is identical on all nodes.
Therefore users can leverage the existing checkpoints to resume in the presence of a failure.

\section{Experiments}
\label{sec:experiments}
In this section, we show the effectiveness of the FanStore design and implementation with both benchmarks and real world applications with analysis and discussion.

\subsection{Hardware and Software Stack}
Throughout the performance evaluation, we use two clusters.
The {\bf GPU Cluster} is Nvidia GTX 1080 Ti based.
It has 24 nodes, each with four 1080 Ti GPUs and $\sim60$~GB local SSD space. 
The nodes are connected by a Mellanox FDR Infiniband interconnect with up to 56Gbps bandwidth and a sub-micro second latency.
The {\bf CPU Cluster} is a 512 node cluster each with two Intel Xeon Platinum 8160 processors and $\sim144$~GB local SSD space.
The interconnect is a 100Gb/sec Intel Omni-Path (OPA) network with a fat tree topology.

On the GPU cluster, DL frameworks use CUDA 9.0 and CUDNN 7.0.
Both clusters run CentOS 7.4, Intel MPI 17.0.3, TensorFlow 1.8.0, TensorLayer~\cite{Dong2017tensorlayer} 1.9.1, Keras~\cite{Chollet2015keras} 2.2.2, and Horovod~\cite{sergeev2018horovod} 0.13.4.

\subsection{Benchmark}
Besides the applications introduced in ~\S\ref{sec:apps}, we design a customized benchmark to measure FanStore's reading performance at scale. 
This benchmark has four file sizes: 128~KB, 512~KB, 2~MB, and 8~MB. 
Each file size has \{128K, 32K, 8K, 2K\} file count, respectively.
At each scale, each node reads all files in the directory, and reports time-to-solution and bandwidth. 
The benchmark reports the aggregated bandwidth and throughput as results.

\subsection{Data Preparation Cost}
In this experiment, we profile the time consumed by data preparation with the three datasets from the real world applications.
Preparing the ImageNet-1k, SRGAN, and RCNN dataset takes 13, 11, and 14 minutes on a single node with Intel Xeon E5-2680 CPU. 
In particular, enabling compression when preparing SRGAN dataset takes 47 minutes, which is a 4.3x slowdown compared to compression-free case.
The cost of data preparation is a one time cost, prepared datasets are used repeatedly for model training.

\subsection{Single Node Performance}
\label{subsec:benchmark}
In this experiment, we compare FanStore's performance against alternative storage options and techniques on a single node with benchmarks and real applications.
The goal is to show FanStore's performance is close to that of the fastest hardware. 

\subsubsection{Benchmark}
\begin{figure*}[h]
	\includegraphics[width=170mm]{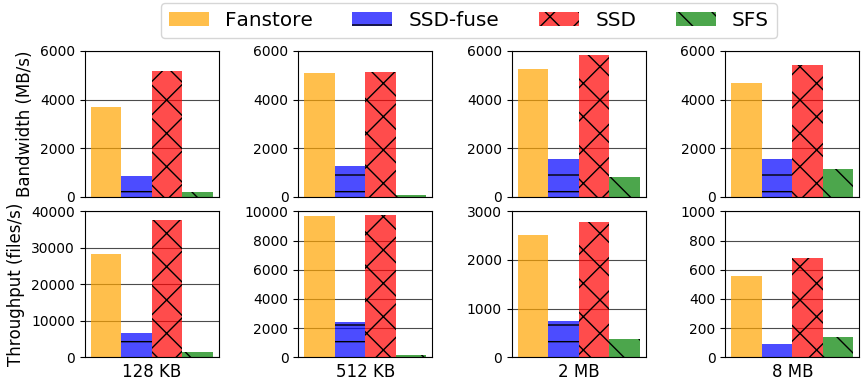}
	\caption{Bandwidth (MB/s) and Throughput (files/s) Comparison with FanStore, Solid State Disk (SSD), Solid-State-Disk via FUSE (SSD-fuse), and Shared File System (SFS) using Benchmark}
 	\label{fig:fs-ssd-fuse}
\end{figure*}

Figure~\ref{fig:fs-ssd-fuse} compares the bandwidth (MB/s) and throughput (files/s) using the benchmark with FanStore, Solid State Disk (SSD), Solid State Disk with FUSE (SSD-fuse), and the Lustre Shared File System (SFS).
In our test deployment, FanStore uses the local SSD as its storage back-end, thus the SSD performance is the upper bound of FanStore's.
As we can see across the scales, FanStore achieves 71\%-99\% of SSD's bandwidth.
The other technical approach of user space file system implementation SSD-fuse is 2.9-4.4X slower as FanStore.
FanStore is also 4.0-64.7x faster than the Lustre shared file system, especially for small files. 

\subsubsection{Applications}
\label{sec:single-apps}
\begin{figure}[h]
	\includegraphics[width=85mm]{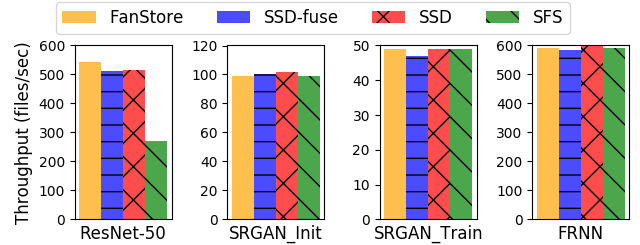}
	\caption{Training Throughput (Items/sec) with Data Stored on Different Hardware}
 	\label{fig:single-node-apps}
\end{figure}

~\zhao{Rerun applications with SSD-fuse}
Figure~\ref{fig:single-node-apps} shows ResNet-50, SRGAN, and FRNN performance with data stored in FanStore, SSD, SSD-fuse, and SFS. 
The application performance is reported as files per second (files/sec), assuming each file contains only one training item.
ResNet-50 runs 5.3\% faster with FanStore than SSD due to directory metadata caching, and it is 2.0x faster compared to the SFS case.  
The sustained throughput of ResNet-50 with FanStore is 544~files/s. 

On the other hand, both the initialization and training stages of SRGAN have identical performance across all options.
This is due to its large amount of computation in each iteration.
Even with the fastest storage in this case, the sustained throughput is 102 and 49~files/s for SRGAN-Init and SRGAN-Train, respectively. 
Similarly to SRGAN, FRNN performs almost identically with four storage options.

\subsection{Multi-node Performance}
In this experiment, we measure the benchmark and real application performance across scales on the GPU cluster and CPU cluster to verify the effectiveness of FanStore's design of scalability. 
If not otherwise specified, each file has only one copy across compute nodes in subsequent experiments.

\subsubsection{Benchmark}
\begin{figure}[h]
	\includegraphics[width=85mm]{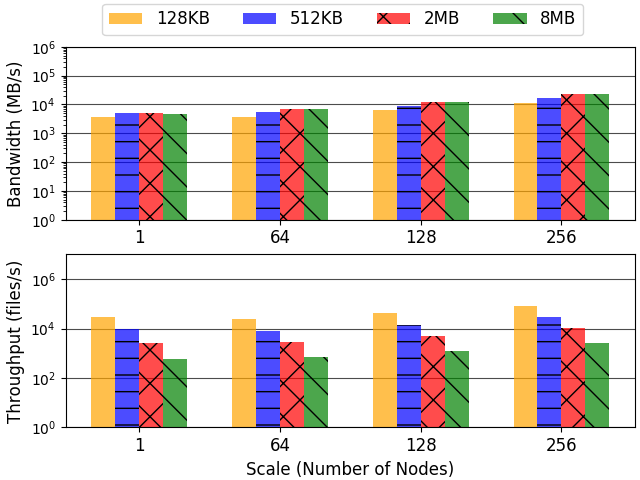}
	\caption{Bandwidth (MB/s) and Throughput (files/sec) Scalability of FanStore on the GPU Cluster}
 	\label{fig:bench-scale-gtx}
\end{figure}

On the GPU cluster, we run the benchmark on \{1, 4, 8, 16\} nodes. 
There are \{4, 16, 32, 64\} GPUs at each scale.
Figure~\ref{fig:bench-scale-gtx} shows the aggregated bandwidth (MB/s) and throughput (files/s) with varying file sizes.
From one node to four nodes, the aggregated bandwidth increases by 1.0-1.5x, with larger file size having larger improvement.
It is not linearly scaling, as the I/O traffic moves from the local storage to network communications across four nodes. 
The aggregated bandwidth scales to 16 nodes with 76.3\%-83.1\% scaling efficiency compared to that of four nodes.
A larger file size produces better scaling performance.
This scaling efficiency compares to the baseline on four nodes, where each node has 25\% of the data stored in local storage.
The local storage hit rate goes down from 25\% to 6.25\% on 16 nodes, which explains the 76.3\%-83.1\% scaling efficiency.

The aggregated throughput of 128KB file size on four nodes achieves only 86.2\% of that of a single node. 
Compared to the single node case, 75\% of the data is accessed through interconnect. 
Given that 128KB file size only utilizes 55.2\% of the peak bandwidth, such performance regression is due to the interconnect latency. 
The aggregated throughput with varying file sizes scales to 16 nodes similarly as with the bandwidth.

\begin{figure}[h]
	\includegraphics[width=85mm]{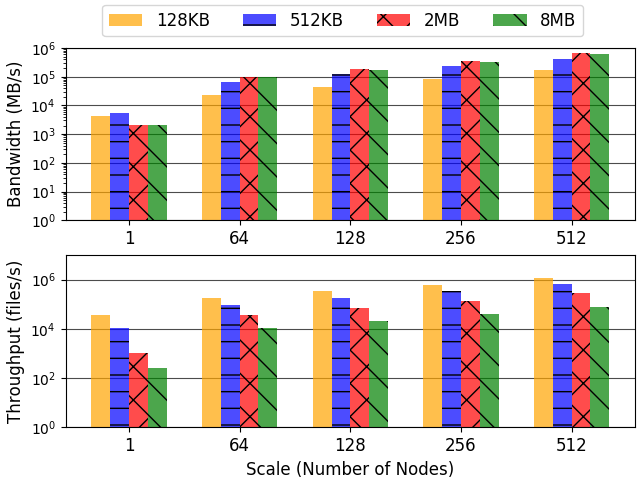}
	\caption{Bandwidth (MB/s) and Throughput (files/sec) Scalability of FanStore on the SKX Cluster}
 	\label{fig:bench-scale-skx}
\end{figure}

To further study FanStore's scaling efficiency, we run the same benchmark on the CPU cluster at the scale of \{1, 64, 128, 256, 512\} nodes, as shown in Figure~\ref{fig:bench-scale-skx}. 
Each node in the CPU cluster has two sockets. 

From one node to 64 nodes, the aggregated bandwidth increases only by 5.8-45.4x across file sizes, with larger file size having better speedup.
The reason is the same as with the GPU cluster, where the I/O is in the form of network communication rather than local storage. 
On 64 nodes, only $1/64$ of the data is local to each node, since we only store one copy of the dataset.
With the bandwidth on 64 nodes as the baseline, FanStore on 512 nodes achieves 81.4\%-88.2\% scaling efficiency, which is better than the measurement on the GPU cluster.
Compared to the GPU cluster scalability, FanStore scales better on the CPU cluster. 
This is because we are comparing against the 64 node case as baseline, with the local storage hit rate of 1.56\%. 
The local storage hit rate goes down from 1.56\% to 0.2\% on 512 nodes, which is much less than the drop from 25\% to 6.25\% in the GPU cluster case.
With most of I/O traffic going through network in the baseline, FanStore shows a better scalability on 512 nodes.

The aggregated throughput has the same pattern as the bandwidth. 
By comparing the throughput and bandwidth, we see that 128~KB and 512~KB file size are network latency-bounded, while 2~MB and 8~MB file sizes are network bandwidth-bounded.

\subsubsection{Applications}
In the previous experiments, we have shown FanStore's scalability up to 512 nodes in the CPU and GPU clusters. 
In this experiment, we study how such scalability applies to real applications.
We run ResNet-50, SRGAN, and FRNN across scales on the GPU and CPU cluster, and present the weak scaling performance with the metric of files per second (files/s). 

\begin{figure}[t]
	\includegraphics[width=85mm]{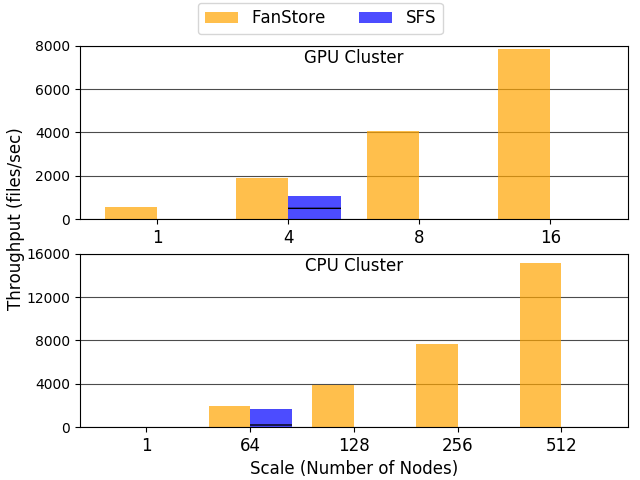}
	\caption{ResNet-50 with ImageNet-1k Scalability Using FanStore on GPU and CPU Clusters}
 	\label{fig:ResNet-50}
\end{figure}

Figure~\ref{fig:ResNet-50} presents ResNet-50 performance across scales on the two clusters. 
We also show ResNet-50's performance with the Lustre shared file system (SFS) using 4 nodes on the GPU cluster and 64 nodes on the CPU cluster. 
Since the testbeds in the experiment are production systems, we were not allowed to run I/O intensive workloads at large scales.

On four nodes in the GPU cluster, ResNet-50 runs 76.1\% faster with FanStore than the shared file system.
And the sustained scaling efficiency is almost 100\% on 16 nodes compared to that on four nodes.
Assuming, the shared file system has linear scalability to 16 nodes, FanStore still runs 68.9\% faster.
In reality, the shared file system can not scale in a linear fashion, and the performance can fluctuate depending on the workload~\cite{xie2012characterizing}.
In contrast, FanStore's performance relies only on the interconnect and local storage. 
It is less prone to be affected by I/O from other jobs on the same cluster.

On 64 nodes in the CPU cluster, ResNet-50 runs 17.1\% faster with FanStore than the shared file system.
The sustained weak scaling efficiency is 95.4\% on 512 nodes compared to that of 64 nodes.

\begin{figure}[h]
	\includegraphics[width=85mm]{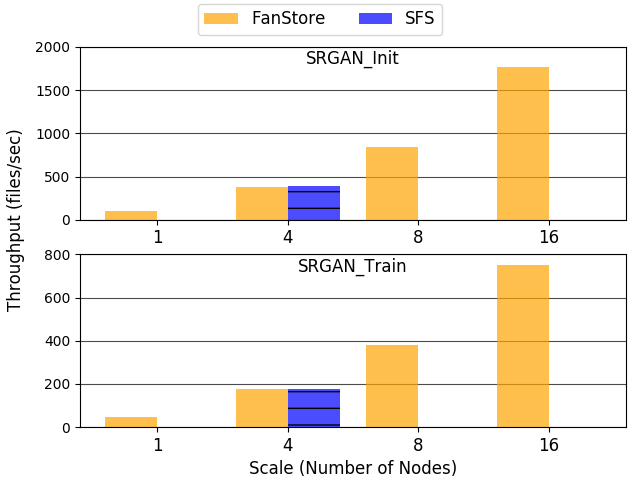}
	\caption{SRGAN Scalability Using FanStore on the GPU Cluster}
 	\label{fig:SRGAN}
\end{figure}

Figure~\ref{fig:SRGAN} shows the scalability of SRGAN\_Init and SRGAN\_Train on the GPU cluster. 
This SRGAN implementation is not feasible on the CPU cluster as some functionality is not implemented in the CPU TensorFlow back end. 
Both stages of SRGAN scale with almost 100\% efficiency from one node to 16 nodes. 
Compared to ResNet-50, this is due to the higher computation requirement in each training iteration of SRGAN.  

\begin{figure}[t]
    \includegraphics[width=85mm]{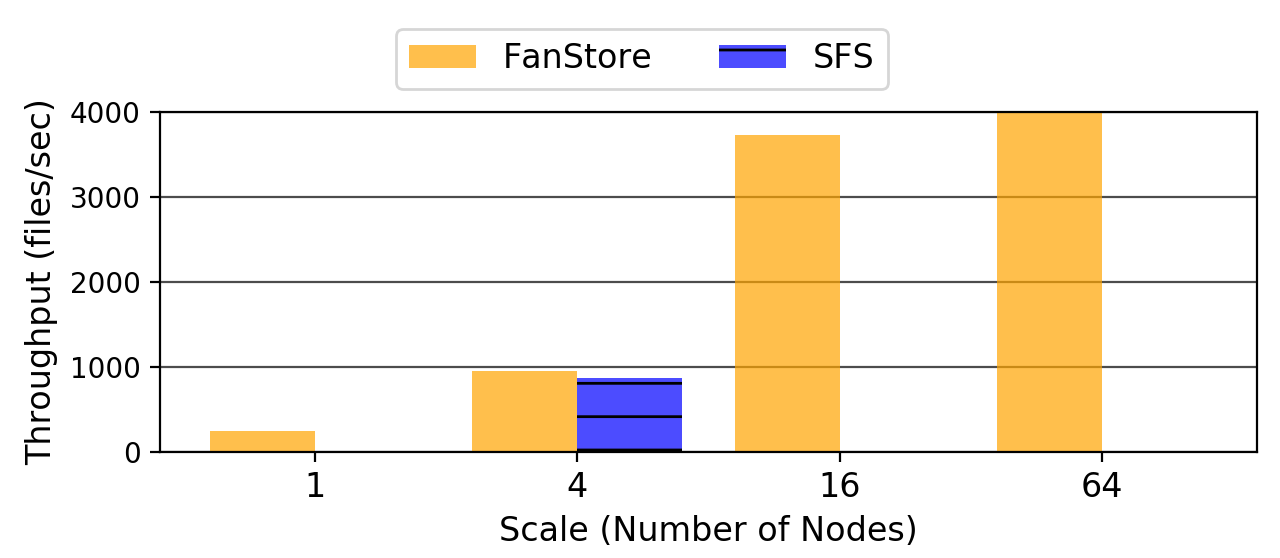}
    \caption{FRNN Scalability Using FanStore on the CPU Cluster}
    \label{fig:FP}
\end{figure}

Figure~\ref{fig:FP} illustrates the performance of FRNN across scales on the CPU cluster. 
Since the complete dataset is $\sim54$~GB and each each local disk has 144~GB free space, so we simply use FanStore's broadcast function to replicate the dataset across all nodes. 
All I/O traffic are completed within the local node. 
As expected, FRNN shows near-linear scalability with 93.1\% efficiency on 64 nodes.
Compared to the shared file system, FanStore runs 9.2\% faster on four nodes.

On both clusters, FanStore enables highly scalable performance across increasing node counts. 
From the shared file system perspective, all runs across different scales read the same dataset.
The preprocessed dataset has a fixed number of files: 48 for the GPU cluster and 512 for the CPU cluster.
These files are loaded to local storage only at the beginning of the training process. 
Thus the I/O workload to/from the shared file system remains constant across different scales of training.

Even though SRGAN and FRNN perform similarly with data stored in FanStore and the Lustre shared file system at small scale,
using FanStore dramatically reduces the I/O workload to the shared file system. E.g, the 0.6 million file inputs during the 200-epoch SRGAN training is now served with 512 large file reads and 120 million network round trip messages.
Such I/O reduction results in less risks for shared file system performance degradation or unresponsiveness.

\subsection{Compression}
FanStore implements compression using the LZSSE8 library. 
In this section, we use both the benchmark and the SRGAN application to verify the effectiveness and study the scalability of this design.

The SRGAN use case has a training dataset of 455~GB and 50076 images in total. 
After compression, the size of the dataset is 163~GB and the compression ratio is 2.8x. 
For the benchmark, we used data sampled from the SRGAN dataset, and generate files with size of \{128~KB, 512~KB, 2~MB, 8~MB\}. 
The compression ratio is roughly identical to the SRGAN dataset.

\begin{figure}[t]
	\includegraphics[width=85mm]{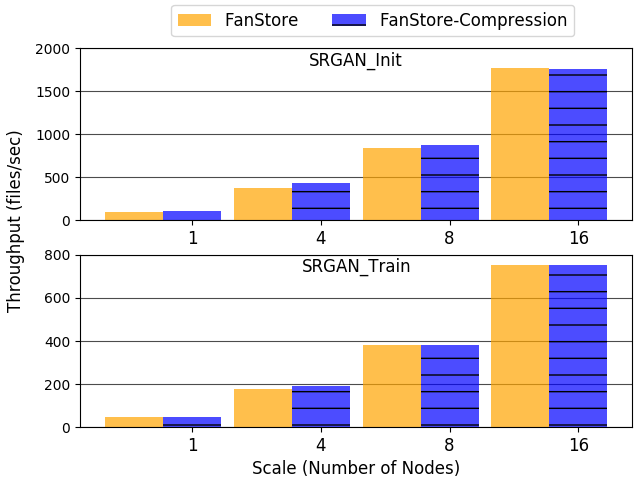}
	\caption{SRGAN Performance with Compressed Data using FanStore on the GPU Cluster}
 	\label{fig:SRGAN-compression}
\end{figure}

We run SRGAN with the compressed dataset on multiple scales of the GPU cluster and compare the throughput (files/sec) to the original FanStore performance. 
Figure~\ref{fig:SRGAN-compression} shows the performance in both cases side by side. 
Both SRGAN stages have an improvement of 2.8\%-11.6\% with compressed data compared to the original case.
This improvement is contributed by the difference between time saved by reading a smaller file and the time to decompress it.

The impact of compression on applications is case by case. 
For example, the ImageNet-1k dataset does not have additional room for compression. 
FanStore leaves compression as a user option, and we encourage users to profile the compression effectiveness before productive training.

By any chance, if the compressed dataset fits in local storage, broadcasting the dataset using FanStore can serve the I/O completely from local storage.

\begin{figure}[t]
	\includegraphics[width=85mm]{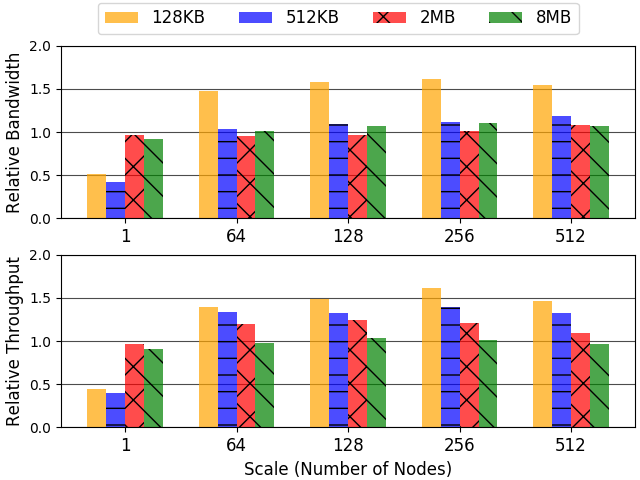}
	\caption{Relative Bandwidth and Throughput with Compressed Data Compared to Uncompressed Data}
 	\label{fig:compression-scale-skx}
\end{figure}

Figure~\ref{fig:compression-scale-skx} presents the relative performance to the compression-free case across scales using the synthetic dataset with 2.8x compression ratio. 

We see that on one node, smaller file sizes bandwidth is only $\sim50$\% of the compression-free case.
This is because the fact of reading small files is throughput bounded, the bound factor is the CPU clock rate in this case, as neither bandwidth nor throughput is at peak.
Consuming more CPU cycles to decompress the data further slows down the performance.
In contrast, the reading performance on larger files, e.g., 2~MB and 8~MB, are close to the compression-free case, as these operations are bounded by local storage bandwidth, and the CPU can use cycles to decompress data.

Overall, using compression in this benchmark results in higher I/O bandwidth and throughput across scales.
Going from one node to 64 nodes, the majority of I/O traffic is going through the high speed interconnect. 
Additionally, the reading operations on smaller files are network latency-bound while the larger files are bandwidth-bound.
We see that reading 128~KB (46~KB after compression) files is bounded by network latency, and
reading files that are larger than 512~KB (183~KB after compression) is bounded by bandwidth.
The sustained scaling efficiency is within the range of 89.2\%-93.5\%.

\subsection{Discussion}
FanStore's capacity of I/O improvement is way beyond ResNet-50, SRGAN, and FRNN.
The average file size of ImageNet-1k is $\sim108$~KB. 
Compared to the benchmark throughput with 128~KB file on the GPU cluster, ResNet-50 only uses about 7.8\%-9.5\% of the sustained peak throughput of FanStore.
ResNet-50 has 50 layers and 1.5 billion single precision floating operations per image.
That indicates FanStore can keep the scaling curve of an application with 10.5-12.8x less computation per image as ResNet-50.
From the processing element's perspective, FanStore can keep the ResNet-50's scalability with 10x more powerful hardware.

Up to 512 nodes on the CPU cluster, FanStore does not hit the ceiling of scalability. 
This is due to the relaxed I/O consistency and the distributed metadata and data service. 
FanStore's performance largely relies on the local storage and interconnect. 
With proper replication setting, FanStore can keep up the scalability with the underlying interconnect.

\section{Conclusion}
\label{sec:conc}
In this paper, we present FanStore, a transient runtime file system that leverages local storage in computer clusters to enable efficient and scalable distributed deep learning training. 
FanStore incorporates the design and techniques of replicated and distributed metadata management, function interception, collective data management, and generic data compression.
FanStore preserves global namespace of the dataset and the POSIX-compliant interface with only user space programs.
Both benchmark and real applications show that FanStore achieves close to hardware limit performance on single node.
And, FanStore scales up to 512 compute nodes with over 90~\% weak scaling efficiency. 
The design and implementation of FanStore dramatically enhances the capability of existing hardware and software in supporting the emerging distributed deep learning applications. 

\section{Future Work}
\label{sec:future}
Our future plan includes collaborating with several research groups to enable more applications, 
reaching out to other data centers and computing centers for deployment opportunities, 
and including non-x86 architectures.
We are also working on the documentation of FanStore and planning to release the source code in late 2018.

%

{\footnotesize \bibliographystyle{acm} \balance
\bibliography{fanstore}}

\theendnotes

\end{document}